\documentclass[aps,pra,showpacs,superscriptaddress]{revtex4-2}  

\usepackage{graphicx}   
\usepackage{bm}             
\usepackage{amssymb}   
\usepackage{amsmath}
\usepackage{color}

\newcommand{\bohr}{$a_0$ }
\newcommand{\npone}{\mbox{$N\!+\!1$}}
\newcommand{\Rmat}{\mbox{$R$-matrix}~}

\begin{document}


\title{Electron-impact cross sections for dissociation processes of vibrationally excited CH radical}

\author{O.~Abidi}
\affiliation{LSAMA, Department of Physics, Faculty of Science of Tunis, University of Tunis El Manar, 2092 Tunis, Tunisia}

\author{I.~Jendoubi}
\affiliation{LSAMA, Department of Physics, Faculty of Science of Tunis, University of Tunis El Manar, 2092 Tunis, Tunisia}

\author{M.~Telmini}
\affiliation{LSAMA, Department of Physics, Faculty of Science of Tunis, University of Tunis El Manar, 2092 Tunis, Tunisia}

\author{R. Ghosh}
\affiliation{Department of Mathematics, Sukumar Sengupta Mahavidyalaya, Keshpur, Paschim Medinipur 721150, India}

\author{K.~Chakrabarti}
\affiliation{Department of Mathematics, Scottish Church College, 1 \&  3 Urquhart Sq., Kolkata 700006,  India}

\author{V.~Laporta}
\email{vincenzo.laporta@cnr.it}
\affiliation{Istituto per la Scienza e Tecnologia dei Plasmi, CNR, Via Amendola, 122/D, Bari, Italy}
\affiliation{Institut de Physique de Rennes, Université de Rennes, 263 avenue du Général Leclerc, Rennes, France}

\begin{abstract}
This paper presents a theoretical investigation of the cross sections for dissociative electron attachment and dissociative excitation processes in vibrationally excited CH radicals induced by electron impact. Resonant electron-CH collisions are analyzed using the \textit{ab-initio} \Rmat method, while nuclear dynamics are explored within the Local Complex Potential framework. A comprehensive set of vibrationally resolved cross sections and rate coefficients is provided for both the ground and first excited electronic states of the CH molecule. These findings contribute to a better understanding of the kinetics of non-equilibrium systems containing CH molecules  with applications in plasma technologies for CO$_2$ reduction, combustion processes and various astrophysical contexts.
\end{abstract}

\pacs{xxxxx}

\maketitle

\section{Introduction \label{sec:intro}}

The methylidyne radical (CH), despite its seemingly simple structure, holds a position of remarkable significance across a diverse array of scientific disciplines, spanning the vast reaches of interstellar space to the forefront of sustainable energy research here on Earth.

In the Interstellar Medium (ISM), CH plays a crucial role in understanding astrophysical processes\,\cite{Morris_2016, Wiesemeyer_AA2018}. In the cold, diffuse expanses of the ISM, where temperatures plummet to tens of Kelvin and densities are exceptionally low, CH emerges as a critical astrophysical probe\,\cite{1994ApJ...424..754A}. Its presence, in fact -- detected through characteristic spectral lines in absorption against background starlight -- provides invaluable insights into the physical and chemical conditions of these regions. Moreover, as a key constituent of cold, dark molecular clouds, CH participates in a complex interplay of gas-phase reactions driven by cosmic rays and ultraviolet radiation\,\cite{1982ApJ...257..125F}. Understanding the formation and destruction pathways of CH in these environments is essential for unraveling the intricate processes of interstellar chemistry and the evolution of molecular complexity in the Universe. The relative abundance of CH in the ISM has long intrigued astrophysicists, particularly as chemical models have struggled to explain this excess\,\cite{annurev:/content/journals/10.1146/annurev-astro-081915-023409}. These spectroscopic and collision studies on CH thus serve as a fundamental basis for further investigations.

Moving beyond astrophysical significance, CH's importance extends to sustainable energy research, particularly in the context of plasma technologies aimed at reducing CO$_2$ emissions\,\cite{LI2020118228}. As the global community grapples with the urgent need to mitigate climate change and transition toward renewable energy sources, CH presents substantial potential in plasma-based technologies\,\cite{C8RA03146K}. By harnessing the energy of plasmas, researchers are attempting to convert CO$_2$, a major greenhouse gas, into valuable products, effectively closing the carbon cycle and paving the way for a more sustainable future\,\cite{Chen:2016aa}. This conversion is not only vital for reducing atmospheric CO$_2$ levels but also for generating hydrocarbons that can serve as renewable energy sources.

Furthermore, CH's relevance extends to combustion processes, particularly in hydrocarbon plasmas, where it significantly influences the formation of final products. CH is an integral component of flames, and its reactivity, particularly in electronically excited states, accelerates combustion\,\cite{LI20101087, Versailles2016QuantitativeCM}. These interactions underscore the necessity of reliable collision data for CH, which is crucial for building chemical kinetic models that accurately describe such reactions. Understanding electron collisions is especially important, as they trigger processes like excitation, dissociation, and dissociative attachment, each of which has profound implications for enhancing the efficiency and selectivity of chemical reactions in plasma systems.

Numerous studies have focused on the spectroscopic properties of CH, such as those reported in the Refs.\,\cite{BEMBENEK19901, 10.1063/1.451140}, which have mapped out its electronic states through \textit{ab initio} quantum chemistry methods\,\cite{Chakrabarti_CH_2019, doi:10.1063/1.2721535, doi:10.1063/1.480285, doi:10.1080/00268976.2014.963183}. Despite these advances, there remains a significant gap in studies on electron collisions involving CH. Total cross sections for elastic scattering, electronic, and rotational excitation, as well as ionization, have been calculated by Joshipura \textit{et al.}\,\cite{JOSHIPURA1997361, Joshipura_2001}, while Baluja and Msezane\,\cite{Baluja_2001} have used \Rmat calculations to derive cross sections for elastic scattering and electronic excitation to CH’s first four excited states. More recently, Ghosh \textit{et al.}\,\cite{Ghosh_2020} expanded on this work, improving previous results by identifying negative ion resonances. These new 
contributions are particularly important for plasma modeling, especially in estimating dissociation and vibrational excitation cross sections, which were not available in previous studies.

In this work, we will focus on calculating of the low-energy cross section for electron collisions with the CH molecule resolved on the vibrational states while we will neglect the rotational contributions. Specifically, we investigate dissociative attachment (DA) and dissociative excitation (DE) processes, which are the key processes to understand the destruction mechanism of CH in non-equilibrium plasma environments. The reactions considered in the paper are listed in Table\,\ref{tab:reactions} and in the following of the paper we will refer to them by the label reported in the last column. As a matter of the fact that low-energy electron-molecule scatterings are dominated by anionic resonances\,\cite{BardsleyWadehra1979}, Table\,\ref{tab:reactions} also reports the main resonant contribution and the activation threshold for each process.

\begin{table}
\centering
\begin{tabular}{lccc}
\hline\hline
~~~~Reaction&~~~~Resonant states~~~~&~~~~Threshold (eV)~~~~&~~~Label~~~~\\
\hline
~~~~$e + \mathrm{CH}(\mathrm{X}\,^2\Pi; v) \to \mathrm{C}^-(^2\mathrm{P}) + \mathrm{H}(^2\mathrm{S})$ & $\mathrm{CH}^-(^1\Sigma^+)$ & 4.10 &  DA1
\\
~~~~$e + \mathrm{CH}(\mathrm{X}\,^2\Pi; v) \to \mathrm{C}(^3\mathrm{P}) + \mathrm{H}^-(^1\mathrm{S})$ & $\mathrm{CH}^-(^3\Pi)$ & 2.76 &  DA2
\\
~~~~$e + \mathrm{CH}(\mathrm{X}\,^2\Pi; v) \to e + \mathrm{C}(^3\mathrm{P}) + \mathrm{H}(^2\mathrm{S})$ & $\mathrm{CH}^-(^1\Sigma^+, ^3\Pi)$ & 3.42 &  DE1
\\
~~~~$e + \mathrm{CH}(a\,^4\Sigma^-; v) \to \mathrm{C}^-(^4\mathrm{S}) + \mathrm{H}(^2\mathrm{S})$ & $\mathrm{CH}^-(^5\Sigma^-)$ & 1.46 & DA3
\\
~~~~$e + \mathrm{CH}(a\,^4\Sigma^-; v) \to e + \mathrm{C}(^3\mathrm{P}) + \mathrm{H}(^2\mathrm{S})$~~~~ & $\mathrm{CH}^-(^5\Sigma^-)$ & 2.67 & DE2
\\
\hline\hline
\end{tabular}
\caption{List of the vibrationally state-resolved reactions considered in the manuscript where $v$ represents the vibrational level of the CH molecule. The resonant state specifies the intermediate anion involved in the process. The threshold of the process refers to the $v=0$ of the corresponding initial electronic state. \label{tab:reactions}}
\end{table}

The paper is organized as follow: Section\,\ref{sec:th_model} presents the theoretical framework used to determine the potential energy curves and to calculate the scattering process. In Section\,\ref{sec:results} we discuss the results and Section\,\ref{sec:conc} provides the conclusions.

\section{Theoretical model \label{sec:th_model}}

\subsection{\Rmat calculations \label{sec:rmat}}

For determining the potential energy curves (PEC) for the bound states of the CH radical and for the CH$^-$ anionic resonant states with the corresponding resonant widths, we have used the \Rmat method\,\cite{Burke2011, Carr2012, Tennyson_2024}. Since the full treatment is available in our earlier work\,\cite{Ghosh_2020}, here we only present a summary.

As a variational method, the \Rmat approach begins by dividing the configuration space into two regions: an inner region, defined as a sphere of radius $a$, and an outer region extending beyond it. This division aims to confine short-range interactions, such as electron correlations and exchange, within the inner region while allowing long-range interactions to be treated separately in the outer region using appropriate mathematical techniques. Within the inner region, the (\npone)-electron wave function representing the target (the CH molecule) and the scattered electron is expressed using a close-coupling expansion:
\begin{equation}\label{eq:cc}
\psi_k^{N+1} = {\cal A} \sum_{i,j} a_{i,j,k} \Phi_i(1,\ldots,N)
F_{i,j}(\npone)+\sum_i b_{i,k}\, \chi_i(1,\ldots,\npone) \,,
\end{equation}
where $\mathcal{A}$ is an antisymmetrization operator, $\Phi_i$ is the $N$-electron wave function of the $i^{th}$ target state, $F_{i,j}$ are continuum orbitals, and $\chi_i$ are two-centre $L^2$ functions constructed by making all $(N+1)$-electrons occupy the target molecular orbitals (MOs) to account for the correlation and polarization of the target in presence of the incoming electron\,\cite{Tennyson_2024}. The inner region wave function is then used to build an energy dependent $\mathbf{R}$ matrix at the boundary of the $R$-matrix sphere, defined by:
\begin{equation}
    \mathbf{R}(r,E) = \mathbf{f}(r) \left[ r\,\mathbf{f}'(r) \right]^{-1} \,,
\end{equation}
where $\mathbf{f}(r)$ is the radial part of $\psi_k^{N+1}$. The $\mathbf{R}$ matrix is then propagated to an asymptotic distance $R_{asy}$ and matched with known asymptotic functions\,\cite{NOBLE1984399}. The matching yields the $K$-matrix from which all resonance parameters are derived.

The diatomic version of the UK molecular \Rmat codes, which uses Slater type orbitals (STOs) to build the target wave functions, were used since STOs are known to give better representation of the target wave function. As the CH target is neutral, the numerical orbitals representing the continuum electron were chosen to be spherical Bessel functions and these were used in a partial wave expansion around the molecular center of mass\,\cite{Ghosh_2020}.

The full details of the CH target calculations are given in our previous work\,\cite{Ghosh_2020} which we do not include here for brevity. The following Complete Active Space Configuration Interaction (CAS-CI) model,
\begin{equation}
(1\sigma)^2 (2-8\sigma,1-3\pi,1\delta)^5\,,\label{eq:rmatrixCAS}
\end{equation}
-- which includes the $1\sigma$ orbital frozen and the CAS defined by $(2-8\sigma,1-3\pi,1\delta)^5$ -- was chosen as this gave the best ground state and excitation energies for the low-lying states considered in our calculation. The model in (\ref{eq:rmatrixCAS}) was subsequently adopted in all bound state and resonance calculations.

Concerning the detection of bound states, an $R$-matrix was constructed on the boundary of the $R$-matrix sphere from the inner region solutions of Eq.\,(\ref{eq:cc}). The $R$-matrix was then propagated in the outer region in a potential which, apart from the Coulomb potential, was given by the diagonal and off diagonal terms of the dipole moments of the CH target states. For bound state calculations the $R$-matrices were propagated to $R_{asy}=40$\,\bohr~and then matched with exponentially decaying functions obtained from a Gailitis expansion\,\cite{NOBLE1984399}. Bound states were then found by searching for negative energy solutions of the inner region Hamiltonian using the searching algorithm of Sarpal \textit{et al.}\,\cite{0953-4075-24-17-006} on a nonlinear, quantum defect based grid\,\cite{Ghosh_2020}.

Calculations were repeated over 20 internuclear distances in the interval $[1,8]\,a_0$ to obtain the potential energy curve for the X\,$^3\Sigma^-$ ground state of CH$^-$ anion and for the low-lying states X\,$^2\Pi$ and $a\,^4\Sigma^-$ for CH radical. The absolute energy of the X\,$^3\Sigma^-$ ground state was found to be $-38.425139$\,Hartree at the CH equilibrium $R_e = 2.1160$\,\bohr~while the electron affinity $E_a$ (vertical value) was found to be 1.0072\,eV which was in reasonably fair agreement with the experimental value $1.238\pm 0.008$\,eV quoted by Kasdan \textit{et al.}\,\cite{KASDAN197578}.

For resonance detection, the \Rmat was propagated up to 70\,\bohr~in the outer region and was then matched to asymptotic functions\,\cite{NOBLE1984399} to obtain a $K$-matrix, and subsequently, the eigenphase sum defined by:
\begin{equation}
    \delta(E) = \sum_i \tan^{-1} (K_{ii})\,,
\end{equation}
where $K_{ii}$ are the diagonal elements of the $K$-matrix. It is known that the eigenphase sum undergoes a characteristic jump of $\pi$ in the neighborhood of a resonance, which causes its second derivative to change sign around the resonance position. To detect resonances, this change in sign of the second derivative of $\delta(E)$ was tracked over a suitable energy grid.

\begin{figure}
\centering
\includegraphics[scale=.4]{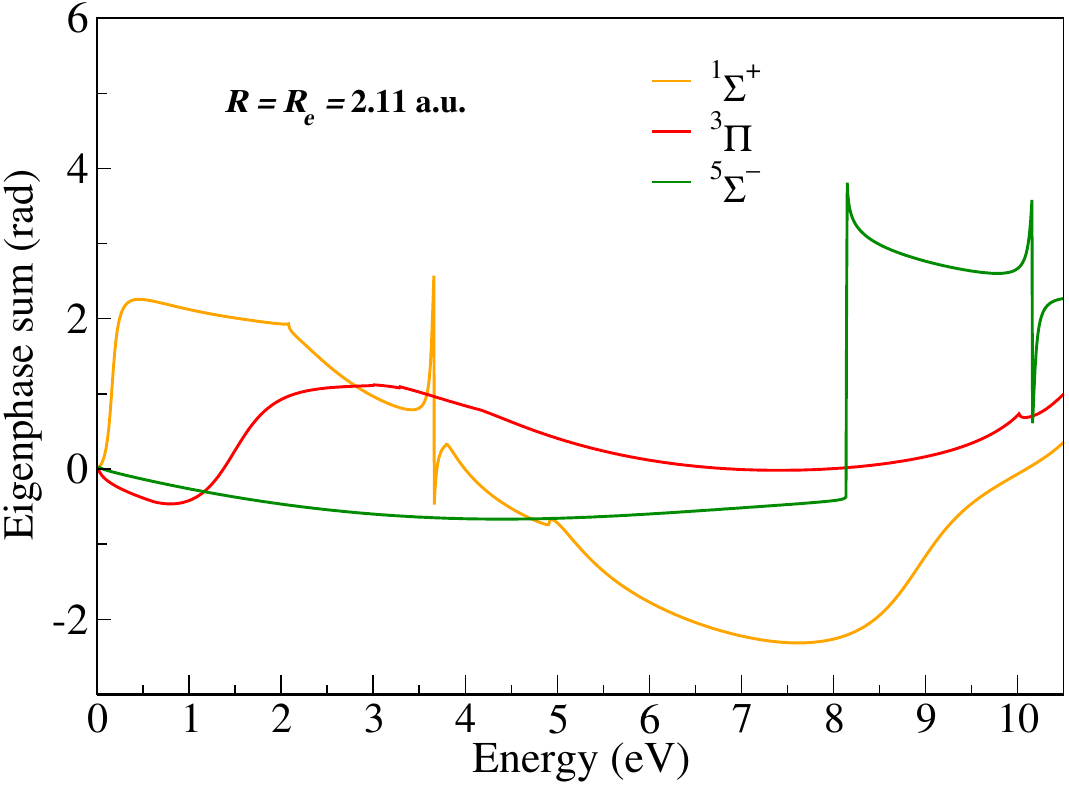} 
\caption{Eigenphase sum for the symmetries $^1\Sigma^+$, $^3\Pi$ and $^5\Sigma^-$ of the CH$^-$ system calculated at CH equilibrium distance. \label{fig:Eigenphase_sum}}
\end{figure}

Figure\,\ref{fig:Eigenphase_sum} presents the eigenphase sum, evaluated at the equilibrium internuclear distance of CH molecule, for the $^1\Sigma^+$, $^3\Pi$ and $^5\Sigma^-$ symmetries of the $e+$CH system up to scattering energy of 10.5\,eV. Concerning the $^1\Sigma^+$ symmetry, two narrow CH$^-$ states are observed: one at very low energy ($E < 1$\,eV) and another one around $E\simeq3.6$\,eV. Since this last resonance is extremely difficult to detect as a function of the internuclear distance, it will be neglected in the following calculations. On the other hand, the $^3\Pi$ symmetry exhibits one broad resonance located approximately at $E \simeq 1.5$\,eV whereas the $^5\Sigma^-$ symmetry reveals a single narrow resonance around $E\simeq 10.1$\,eV. Notably, the discontinuity near 8\,eV, for the $^5\Sigma^-$ symmetry, does not correspond to a typical structure of the resonance and may be attributed to numerical artifact.

Once the resonance positions $E_i^r$ was accurately located, the resonance widths $\Gamma_i^r$ was determined by fitting to a Breit Wigner form\,\cite{Tennyson1984421},
\begin{equation}
    \delta(E) = \sum_i \tan^{-1}\left[ \frac{\Gamma_i^{r}}{E-E_i^r} \right] + b(E)\,,
\end{equation}
where $b(E)$ is the background, which is usually chosen to be a linear or a quadratic polynomial to represent the underlying trend of $\delta(E)$ near the resonance. Table\,\ref{tab:res_pos} reports the list of the low-lying resonances detected at equilibrium distance of CH\,\cite{Ghosh_2020}.
\begin{table}
\centering
\begin{tabular}{ccc}
\hline\hline
~~~Symmetries~~~ & ~~~$E^r$ (eV)~~~ & ~~~$\Gamma^r$ (eV)~~~  \\
\hline
$^1\Sigma^+$ & 0.312  & 0.219 \\
$^3\Pi$            & 1.518  & 0.814 \\
$^5\Sigma^-$  & 10.132   & 0.070 \\
\hline\hline
\end{tabular}
\caption{Low-lying resonance positions $E^r$ and widths $\Gamma^r$ relative to the CH equilibrium distance as obtained from \Rmat calculations\,\cite{Ghosh_2020}.}  \label{tab:res_pos}
\end{table}

Similar to the case of bound states, the resonance calculations were then repeated over 20 internuclear distances obtaining the PEC for three resonant states with symmetries $^1\Sigma^+$, $^3\Pi$ and $^5\Sigma^-$ and the corresponding resonance widths. For the three detected resonances CH$^-$, we found that both symmetries  $^1\Sigma^+$ and  $^3\Pi$ couple with only the parent X\,$^2\Pi$ state of CH whereas the $^5\Sigma^-$ symmetry couples only with the $a\,^4\Sigma^-$ excited state of CH radical.

\subsection{Potential energy curves construction}
We observed the energy of the PEC at large internuclear distances of the low-lying states X\,$^2\Pi$ and $a\,^4\Sigma^-$ (going to the C($^3$P)+H($^2$S) limit) of CH obtained from \Rmat calculations slightly deviated from the values available from the NIST database\,\cite{NIST_ASD}. For this reason, we chose to determine the PEC for the ground and for the first excited state of the CH radical by using \textit{ab-initio} quantum chemistry MOLPRO\,\cite{doi:10.1063/5.0005081} package which have better agreement with NIST data.

The electronic structure calculations were performed using MOLPRO’s state-averaged CASSCF/MRCI methods. The active space included five orbitals occupied by seven electrons: the carbon atom’s $2s$, $2p_x$, $2p_y$, and $2p_z$ orbitals, as well as the hydrogen atom’s $1s$ orbital. This choice ensures a balanced description of valence electron correlation effects, particularly for the ground state X\,$^2\Pi$ and the excited state $a\,^4\Sigma^-$ of the CH radical.

Dynamic electron correlation was incorporated through MRCI calculations with single and double excitations, applying the Davidson (+Q) correction\,\cite{DAVIDSON197587} which corrects for size-extensivity in MRCI by estimating the effects of unlinked quadruple excitations. This choice aligns with standard practices for improving dynamic correlation treatment in multi-reference methods. The aug-cc-pVQZ basis set\,\cite{doi:10.1021/acs.jcim.9b00725} was employed for both carbon and hydrogen atoms.

In addition to the two CH target electronic states (X\,$^2\Pi$ and $a\,^4\Sigma^-$), bound regions of the $^3\Pi$ and $^5\Sigma^-$ resonant states were also computed beyond their crossing points with the parent state using MRCI approach.

The next step involved has been the matching between MRCI results with the \Rmat calculations. We adopted a straightforward approach: initially, we retained the PEC configurations for CH and CH$^-$ as obtained from MRCI method. Next, we aligned the $^1\Sigma^+$ resonant PEC from the \Rmat calculations by shifting it to match the asymptotic energy threshold relative to neutral CH. Subsequently, we adjusted the resonant $^3\Pi$ and $^5\Sigma^-$ branches from the \Rmat results to smoothly connect with the corresponding MRCI branches.

\begin{figure}
\centering
\includegraphics[scale=.8]{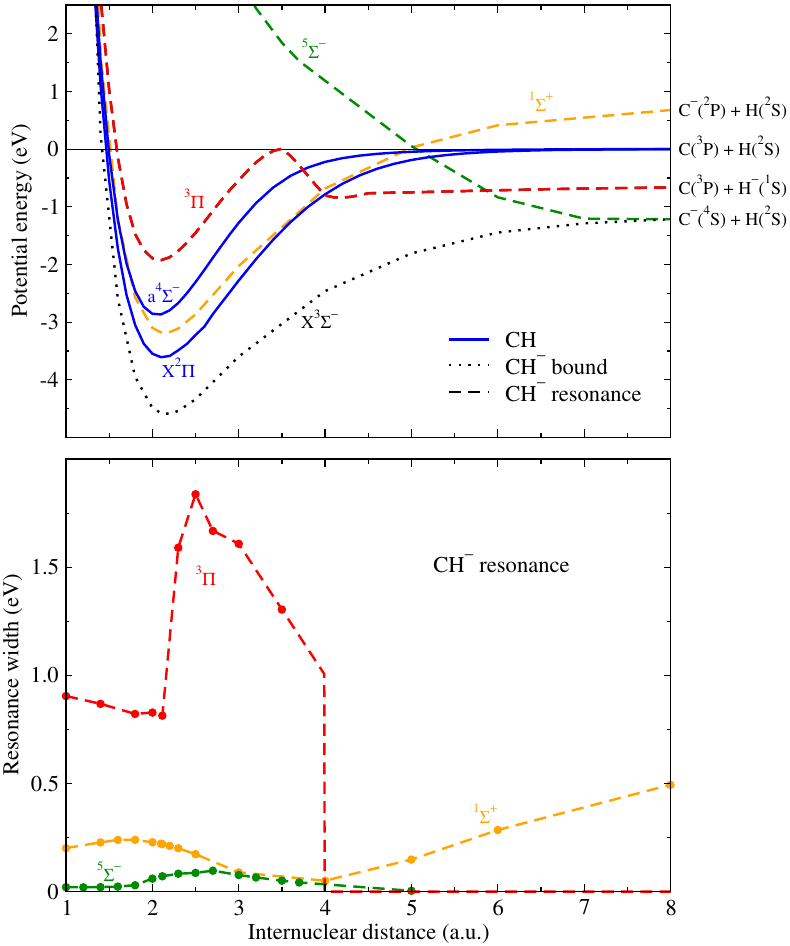} 
\caption{(Top) Potential energy curves for the two low-lying electronic states of CH radical (solid blue lines) as obtained by MRCI approach and for the CH$^-$ anionic states (broken lines) as obtained from \Rmat calculations. (Bottom) Widths of the CH$^-$ resonances considered in the text. Filled circles represent the calculated \Rmat results. \label{fig:CHpot}}
\end{figure}

By applying this procedure, we obtained our final results for the PEC of CH radical and CH$^-$ anionic states summarized in Figure\,\ref{fig:CHpot} with the resonance widths as a function of the internuclear distances. With respect to the reactions in Table \ref{tab:reactions}, in the following of the paper, the electronic states of CH radical and CH$^-$ resonances will be represented by the potentials $V^s(R)$, $s\in{\cal S} = \{\mathrm{X}\,^2\Pi, a\,^4\Sigma^-\}\equiv\{0,1\}$, and $V^-_r(R)$, $r\in{\cal R}  = \{^1\Sigma^+, ^3\Pi, ^5\Sigma^-\}\equiv\{1,2,3\}$ respectively. Table\,\ref{tab:CH-_limits} reports the asymptotic electron affinity for the anionic states of CH$^-$, calculated with respect to the dissociation energy of the target CH molecule, compared with the experimental value available in literature\,\cite{PhysRevA.43.6104, Feldmann1977InfraredPM, PhysRevA.93.013414}.

\begin{table}
\centering
\begin{tabular}{cccl}
\hline\hline
~~~Symmetries~~~ & Limit & \multicolumn{2}{l}{Energy (eV)}  \\
\hline
$^1\Sigma^+$ & $\mathrm{C}^-(^2\mathrm{P}) + \mathrm{H}(^2\mathrm{S})$ & +0.67 &
\\
$^3\Pi$ & $\mathrm{C}(^3\mathrm{P}) + \mathrm{H}^-(^1\mathrm{S})$ & --0.66 & (--0.75 \cite{PhysRevA.43.6104}) 
\\
$\mathrm{X}\,^3\Sigma^-$, $^5\Sigma^-$ & ~~~$\mathrm{C}^-(^4\mathrm{S}) + \mathrm{H}(^2\mathrm{S})$~~~ & --1.21 & (--1.26 \cite{Feldmann1977InfraredPM, PhysRevA.93.013414})~~~ 
\\
\hline\hline
\end{tabular}
\caption{Asymptotic electron affinity for the CH$^-$ anionic states, with respect to the dissociation energy of neutral CH molecule, as obtained from MRCI method and \Rmat calculations. Experimental energy values where available in literature are given in brackets for comparison.  \label{tab:CH-_limits}}
\end{table}

Our results for the CH radical PECs are in agreement with those reported by Kalemos \textit{et al.} in Ref.\,\cite{doi:10.1063/1.480285}. Additionally, the resonant states we obtained are consistent with the calculations by Sun \textit{et al.} shown in Fig.\,1 of Ref.\,\cite{doi:10.1063/1.442853}. Specifically regarding the resonance widths: in adiabatic fixed-nuclei calculations, diabatic states manifest as resonances above the target state and therefore exhibit finite widths. Below the target state, the resonance becomes bound, leading to a width of zero. Consequently, the widths of the $^3\Pi$ and $^5\Sigma^-$ resonances decrease sharply around $\sim4$\,$a_0$ and $\sim5$\,$a_0$, respectively. In particular, for the $^3\Pi$ state, the matching between MRCI and \Rmat calculations causes the abrupt decrease of the corresponding width at $\sim4$\,$a_0$. On the other hand, we do not have a physical explication for the observed discontinuity in the $^3\Pi$ width near $\sim2$\,$a_0$. This jump in the width could be caused by a complicated interplay between the orbitals of high excited states. However, non-smooth behavior for the resonant widths has been found also in other systems like ArH$^+$\,\cite{doi:10.1093/mnras/sty1549} and BeH$^+$\,\cite{0741-3335-59-4-045008}.

In order to find the vibrational levels for the potential energy $V^s(R)$ for the X\,$^2\Pi$ and $a\,^4\Sigma^-$ states of CH radical, we applied the Fourier Grid Hamiltonian (FGH) numerical method\,\cite{10.1063/1.456888} to the following Schr\"{o}dinger equation:
\begin{equation}
\left[ -\frac{\hslash^2}{2\mu}\frac{d^2}{dR^2} + V^s(R)  \right]\chi^s_i(R) = \epsilon^s_i\,\chi^s_i(R)\,,
\hspace{1cm} s\in {\cal S} \,, \label{eq:targ_wf}
\end{equation}
where $\mu$ is the reduced mass of CH molecule.

The FGH method involves discretizing Eq.\,(\ref{eq:targ_wf}) by sampling both the kinetic energy operator and the potential $V^s(R)$ over the internuclear distance range $R \in [1,8]\,a_0$. This range is represented as a one-dimensional grid with a step size of 0.001\,$a_0$. The method simultaneously determines the bound vibrational wave functions $\chi^s_{i=v}(R)$ along with their corresponding vibrational energies $\epsilon^s_{i=v}$, as well as the discretized continuum wave functions $\chi^s_{i=c}(R)$ beyond the dissociation threshold.

Applying the FGH method to the X\,$^2\Pi$ and $a\,^4\Sigma^-$ electronic states of the CH molecule, we identified 17 and 12 vibrational levels, respectively. Additionally, we obtained 75 and 80 discretized continuum states extending up to 15 eV. The numerical values of the vibrational levels, along with relevant spectroscopic properties, are presented in Table\,\ref{tab:CHviblev}.
\begin{table}
\centering
\begin{tabular}{ccc}
\hline\hline
& ~~~~$\mathrm{CH}(\mathrm{X}\,^2\Pi)$~~~~ & ~~~~$\mathrm{CH}(a\,^4\Sigma^-)$~~~~\\
\hline
$\mu$ (a.u.) & \multicolumn{2}{c}{1695.2}\\
$R_e$ (a.u.) & 2.12 (2.11) &  2.06 (2.05)\\
$D_e$ (eV)  & 3.61 (3.61) & 2.87 (2.86)\\
$D_0$ (eV) & 3.42 & 2.67\\
\hline
~~~$v$~~~ & $\epsilon^0_{v}$~(eV) & $\epsilon^1_{v}$~(eV)\\
\hline
  0 &   0.00 &    0.00 \\
  1 &   0.33 &    0.36 \\
  2 &   0.65 &    0.71 \\
  3 &   0.96 &    1.04 \\
  4 &   1.25 &    1.34 \\
  5 &   1.52 &    1.63 \\
  6 &   1.78 &    1.89 \\
  7 &   2.03 &    2.12 \\
  8 &   2.26 &    2.32 \\
  9 &   2.47 &    2.48 \\
10  &  2.66 &   2.59\\
11  &  2.84 &   2.65\\
12  &  3.00 \\
13  &  3.14 \\
14  &  3.25 \\
15  &  3.34 \\
16  &  3.40 \\
\hline\hline
\end{tabular}
\caption{Some calculated spectroscopic constants (in braket comparison with the values obtained in the papers\,\cite{doi:10.1063/1.480285, doi:10.1063/1.2721535}) for two low-lying electronic states of CH radical considered in the manuscript and the corresponding energies of the vibrational levels. \label{tab:CHviblev}}
\end{table}

\subsection{LCP model for the nuclear motion \label{sec:LCPmodel}}

In this section, the theoretical framework used to describe the resonant reactions reported in Table\,\ref{tab:reactions} will be briefly presented. The main feature of a resonant collision is that the incident electron is temporarily captured by the molecular target, forming a unstable anionic system  -- the resonance --, and after a characteristic lifetime, related to the resonance width, the negative compound decays, leading to a large spectrum of different final states, competing with each other, including excitation or dissociation. 

Resonant processes can be described at various levels of approximation, some of which include the treatment of nuclear motion. In this study, we will focus specifically on vibrational effects, while neglecting rotational contributions.

The local complex potential (LCP) model, initially developed by Bardsley, Herzenberg and Mandl\,\cite{0034-4885-31-2-302, Domcke199197, PhysRevA.20.194},  is an effective quantum \textit{ab-initio} approach which is able to calculate the resonant scattering taking into account the molecular nuclear dynamics and thus it is able to consider vibrational state resolved cross sections. Recently, the LCP model within \Rmat calculations has been widely used to determine low-energy dissociations by electron impact and vibrational excitations of the molecules of OH\,\cite{Chakrabarti_2019}, NO\,\cite{Laporta_2022, Laporta_2020, 10.1088/1361-6595/ab86d8} and D$_2$\,\cite{Laporta_2021}, which gave results in good agreement with experiments.

According to the LCP approach, the unpolarized cross sections for the DA and DE processes, in the rest frame of CH molecule initially in vibrational level $v$ of the electronic state $s  \in {\cal S}$ and for an incident electron of energy $\epsilon$, are given by:
\begin{eqnarray}
\sigma^{\mathrm{DA}}_{s,v}(\epsilon) &=& \sum_{r\in {\cal R}} \frac{2S_r+1}{(2S_s+1)\,2} \frac{g_r}{g_s} 
8\pi^3\,\frac{K_r}{\mu}\,\frac{m}{k}\,\lim_{R\to\infty}\left|\xi^r_{s,v}(R)\right|^2\,, \hspace{1cm} s \in {\cal S}  \,,  \label{eq:DAxsec}
\\
\sigma^{\mathrm{DE}}_{s,v}(\epsilon) &=& \sum_{r \in {\cal R}} \frac{2S_r+1}{(2S_s+1)\,2} \frac{g_r}{g_s} 
\frac{64\,\pi^5\,m^2}{\hslash^4} 
\int_{D_e}^{\epsilon^{max}} d\epsilon_c \frac{k'}{k} \left | \langle 
\chi_c^{s} | \mathcal{V}_r | \xi^r_{s,v} \rangle \right |^2\,, \hspace{1cm} s \in {\cal S} \,, \label{eq:DExsec}
\end{eqnarray}
where $2S_r+1$ and $2S_s+1$ account for the spin-multiplicities of the resonant anion state and of the neutral target state respectively, $g_r$ and $g_s$ represent the corresponding degeneracy factors, $m$ is the electron mass, $\mu$ is the reduced mass of CH, and $k$ represents the incoming electron momentum.

For DE cross section, $k'$ is the outgoing electron momentum, $\chi^{s}_c$ refers to the final wave function belonging to the continuum part of the electronic state $s$ of CH and parenthesis $\langle \cdot  | \cdot | \cdot \rangle$ stands for integration over the internuclear coordinate $R$. The integral in the DE cross section extends into the continuum part of the CH potential, \textit{i.e.} from the molecular dissociation energy $D_e$ up to $\epsilon^{max} =15$\,eV.

For the DA process, the limit $R\to\infty$ is calculated by taking an average of the resonant wave function over the last 10 points of the grid at $R=8\,a_0$. $K_r$ is the asymptotic momentum of the final dissociating fragments, in our case $\mathrm{C} + \mathrm{H}^-$ or $\mathrm{C}^- + \mathrm{H}$, for the total scattering energy $E=\epsilon^s_v+\epsilon$, given by:
\begin{equation}
K_r^2(E) = \frac{2\mu}{\hslash^2} \lim_{R\to\infty}(E-V^-_r(R))  \,, \hspace{1cm} r\in {\cal R}\,.
\end{equation}
The threshold energy for the DE and DA processes are defined by $V_s(R=8\,a_0) - \epsilon_v^s$ and $V^-_r(R=8\,a_0) - \epsilon_v^s$ respectively.

In Eqs.\,(\ref{eq:DAxsec}) and (\ref{eq:DExsec}), $\xi^r_{s,v}(R)$ represents the resonant wave function solution of the nuclear equation with total energy $E$:
\begin{equation}
\left[ -\frac{\hslash^2}{2\mu}\frac{d^2}{dR^2} + V_r^-(R)  - 
\frac{i}{2}\Gamma^r_s(R) - E \right]\xi^r_{s,v}(R) = -\mathcal{V}^r_s(R)\,\chi^s_v(R)\,, 
\hspace{1cm} s\in {\cal S}  \,,  r\in {\cal R}\,, \label{eq:res_wf}
\end{equation}
where $V_r^-(R)$ with $r=1,2,3$ is the potential energy for the three CH$^-$ resonant states reported on the top of Fig.\,\ref{fig:CHpot} (dashed lines) whereas $\Gamma^r_s(R)$ refers to the autoionization partial widths of the resonance $r$ with respect to the neutral state $s$. As discussed at the end of Section\,\ref{sec:rmat}, the allowed couplings are $\Gamma^1_0(R)\equiv\Gamma^{^1\Sigma^+}$, $\Gamma^2_0(R)\equiv\Gamma^{^3\Pi}$ and $\Gamma^3_1(R)\equiv\Gamma^{^5\Sigma^-}$ showed on the bottom of the Fig.\,\ref{fig:CHpot} whereas $\Gamma^1_1(R)=\Gamma^2_1(R)=\Gamma^3_0(R)=0$. Being the three anion states of different spin multiplicities, therefore they are not coupled, allowing one to solve Eq.\,(\ref{eq:res_wf}) independently for each state.

In the nuclear equation (\ref{eq:res_wf}), $\chi^s_v(R)$ represents the wave function for the initial vibrational state of electronic state $s$ of CH with energy $\epsilon^s_v$ calculated from Eq.\,(\ref{eq:targ_wf}) and $\mathcal{V}^r_s(R)$ is the discrete-to-continuum coupling between the resonance $r$ and the electronic state $s$ of target defined by:
\begin{equation}
{\mathcal{V}^r_s}^2(R) = \frac{\hslash}{8\pi^2} \frac{\Gamma^r_s(R)}{\sqrt{2m\left(V^-_r(R) - V^s(R)\right)}}\,,\hspace{1cm} s\in {\cal S}  \,, r\in {\cal R} \,. \label{eq:couplingVr}
\end{equation}
We solved for the nuclear dynamics by writing in a matrix form the Eq.\,(\ref{eq:res_wf}) and applying the FGH method over the same discretized grid of internuclear distance $[1,8]\,a_0$ used to solve Eq.\,(\ref{eq:targ_wf}).

\begin{figure}
\centering
\includegraphics[scale=.8]{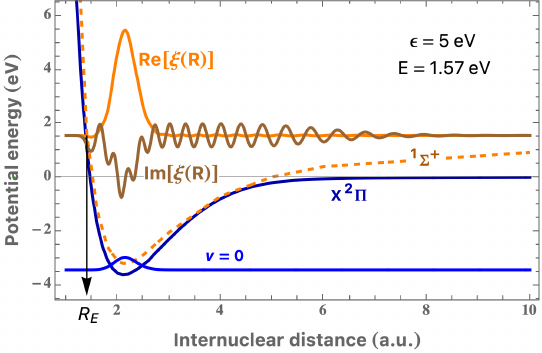} \hspace{.5cm} \includegraphics[scale=.8]{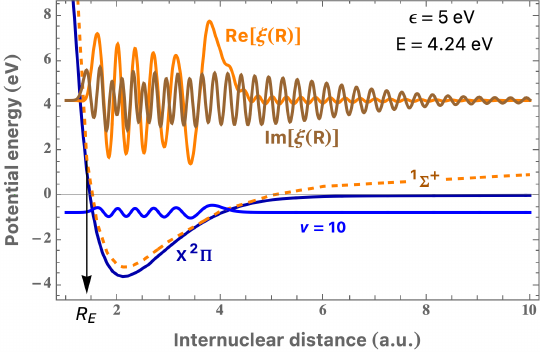} 
\caption{Real and imaginary part of the resonant wave function $\xi^1_{0,v}(R)$ for a collision of CH$(\mathrm{X}\,^2\Pi, v=0)$ (plot on the left) and CH$(\mathrm{X}\,^2\Pi, v=10)$ (plot on the right) with an electron of energy $\epsilon=5$\,eV. Initial vibrational wave functions of CH molecule and the classical turning point $R_E$ are also shown. Amplitudes are not to scale.  \label{fig:reswf_xi}}
\end{figure}

In order to explain as the LCP model works, we show in Fig.\,\ref{fig:reswf_xi} the behavior of the resonant wave function $\xi^1_{0,v}(R)$ (real and imaginary part) compared with the initial vibrational wave function $\chi_v(R)$ of CH radical. The figure refers to the specific case of $s = \mathrm{X}\,^2\Pi$, $r=\,^1\Sigma^+$, incident electron energy $\epsilon=5$\,eV, and for two initial vibrational levels $v=0$ and $v=10$. This configuration, in particular, is suitable to calculate cross section of the process labeled DA1 in Table\,\ref{tab:reactions}. We stress the fact as the amplitudes of the wave functions in Fig.\,\ref{fig:reswf_xi} are not to scale, as they were multiplied by an artificial factor in order to be included correctly in the same graph, but the wave functions were placed at the right position with respect to the potential energies and to the scattering total energy $E$.

First of all, we note as the resonant wave function is strongly influenced by the initial vibrational state $\chi_v(R)$. In fact, the real part of $\xi^1_{0,v}(R)$ function follows, basically, the shape of the CH vibrational wave function at small internuclear distances even if there are also small oscillations at large $R$ (not visible in the graph for the case $v=0$). Imaginary part, on the contrary, shows an oscillating behavior overall range of $R$ with some distortions at small $R$ due to the presence of the initial vibrational states. Specifically, the behavior of the imaginary part of $\xi^1_{0,v}(R)$ drives the shape of cross sections -- where sometimes oscillating structures are present -- in particular for DA processes, as we will show in the next section.

Moreover, we note as for higher vibrational states, the nuclear wave function has a significant amplitude at $R=8\,a_0$, as can be seen in the right panel of Fig.\,\ref{fig:reswf_xi}. In order to check the grid size effect on the resonant wave function and consequently to avoid introducing serious artificial reflections that would affect the calculated cross sections, we extrapolate linearly the potentials determined in Sec.\,\ref{sec:rmat} to $R=10\,a_0$ and we made calculations of resonant wave function and the corresponding cross sections. In the extended range of the grid, we note as the imaginary part of resonant wave function goes to zero correctly and we verified as the corresponding cross sections at with grid size limited to $R=8\,a_0$ and $R=10\,a_0$ do not produce significant effects.

\section{Results and discussion \label{sec:results}}

\begin{figure}
\centering
\includegraphics[scale=.6]{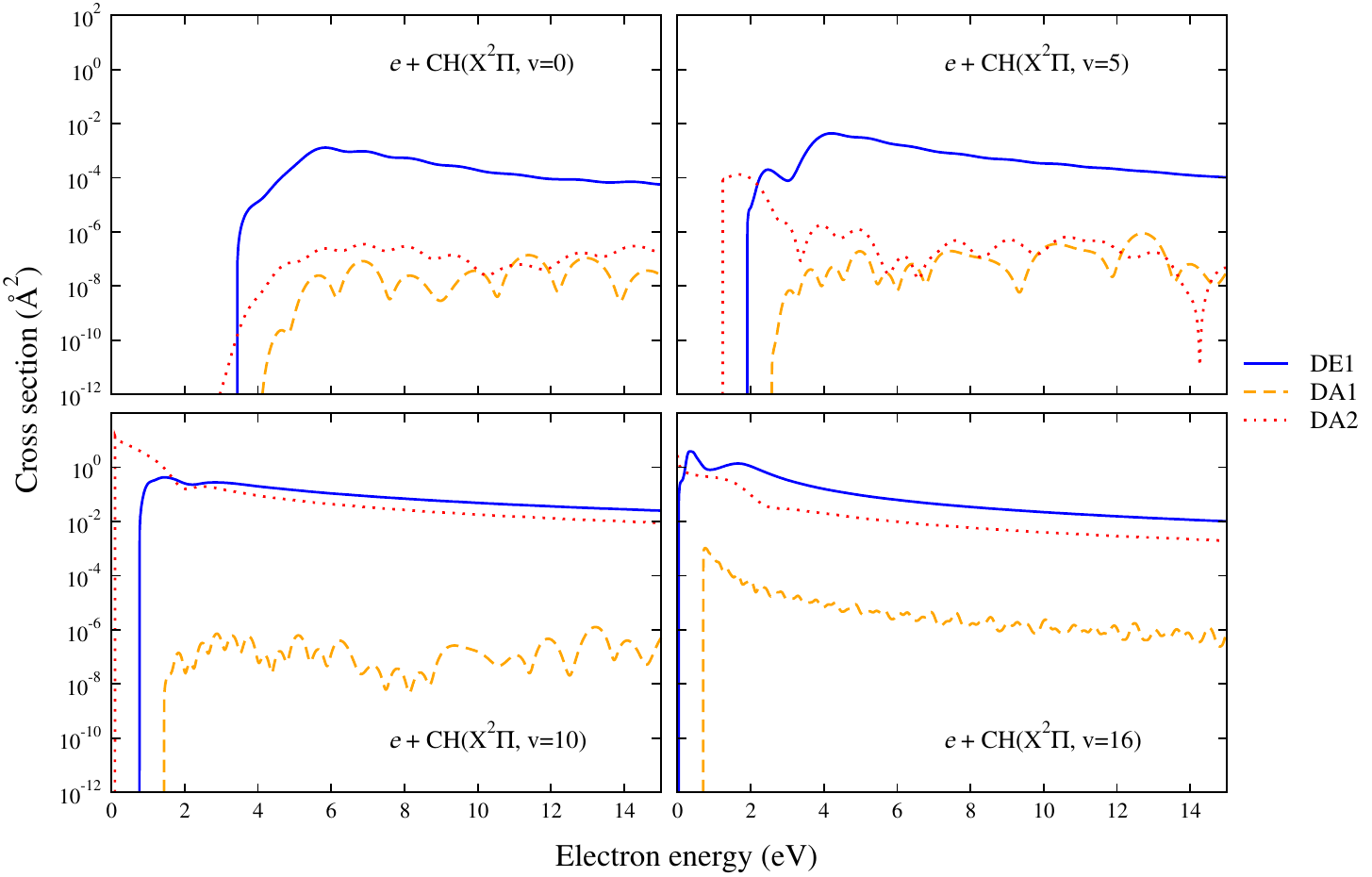} 
\caption{Cross sections for the DA and DE processes from the ground state $\mathrm{X}\,^2\Pi$ of CH molecule for the vibrational levels $v=0,5,10,16$.  \label{fig:CHxsec_X}}
\end{figure}

\begin{figure}
\centering
\includegraphics[scale=.6]{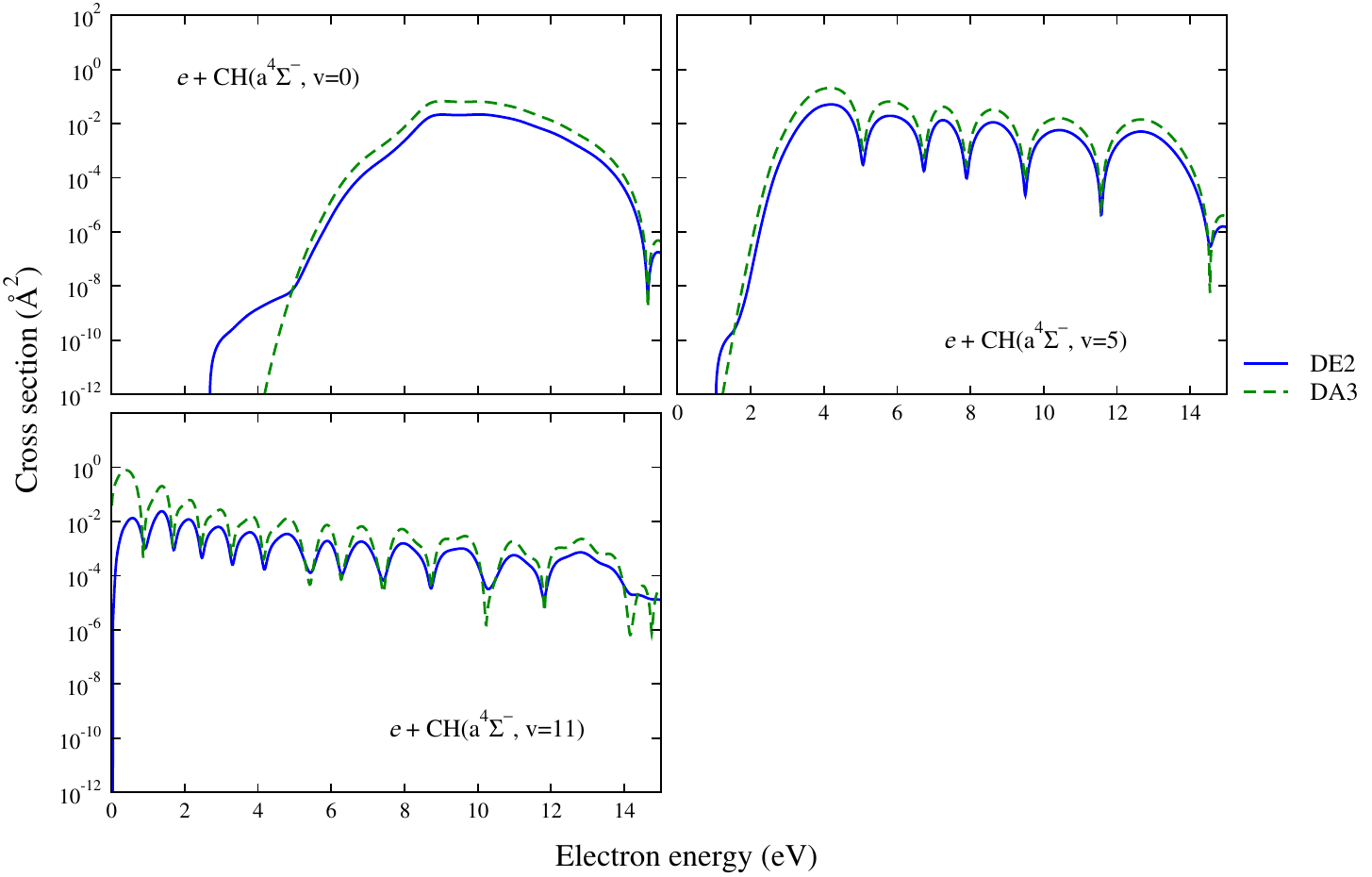} 
\caption{Cross sections for the DA and DE processes from the excited state $a\,^4\Sigma^-$ of CH molecule for the vibrational levels  $v=0,5,11$.  \label{fig:CHxsec_4a}}
\end{figure}

Figures\,\ref{fig:CHxsec_X} and \ref{fig:CHxsec_4a} summarize the main results of this paper, presenting the cross sections for the reactions presented in Table\,\ref{tab:reactions}. These cross sections are resolved over the vibrational level of the CH molecule for the two electronic states, $\mathrm{X}\,^2\Pi$ and $a\,^4\Sigma^-$, as a function of the incident electron energy.

At first glance, the main characteristic we can observe in the cross sections is the presence of oscillatory structures. Such structures are typical of resonant scattering, in particular for dissociative attachment processes, and they have been detected in various other systems, including F$_2$\,\cite{10.1063/1.435896}, HCl\,\cite{10.1063/1.444764}, O$_2$\,\cite{PhysRevA.91.012701} and CO\,\cite{0963-0252-25-1-01LT04}. Since DA process is a fully non-adiabatic process, the precise nature of these oscillations is not yet completely understood. As demonstrated at the end of Section IIC, this oscillatory behavior could be attributed to the shape of the resonant wave function at total energy $E$ and its interplay with the target vibrational wave function. More specifically, it depends on the overlap integral $\left\langle\chi_v^s|\xi^r_{s,v}\right\rangle$  in the Franck-Condon region.

The qualitative features of the cross sections can be better understood within O'Malley's treatment of the DA process\,\cite{PhysRev.150.14, PhysRev.155.59}.  According to the O'Malley approach, the cross section in Eq.\,(\ref{eq:DAxsec}), aside form kinematics factors, can be approximate by the following expression:
\begin{equation}
\sigma^{DA} \propto \left|   \left\langle\chi_v^s|\xi^r_{s,v}\right\rangle   \right|^2 e^{-\rho}\,, \label{eq:DAomalley}
\end{equation}
where $\rho$ is the classical survival factor,
\begin{equation}
\rho = \int_{R_E}^{R_c} \frac{\Gamma^r_s(R)}{\hslash\, v(R)} dR \,. \label{eq:survival}
\end{equation}
The integration limits in the integral extend, over the internuclear distance, from the classical turning point $R_E$ to the stabilization point $R_c$. Specifically, the classical turning point is defined by the condition $V_r^-(R_E)=E$, while $R_c$ is the crossing point between the target and resonant potentials, \textit{i.e.} $V^s(R_c) = V^-_r(R_c)$. The resonant potential curve is considered relatively stable against auto-ionization for $R>R_c$. Inside the integral, $v(R)$ represents the classical velocity of the dissociating atomic fragments.

As we will show in the following discussion, the key quantity governing the shape of a specific set of DA cross sections is the total energy collision, $E=\epsilon+\epsilon^s_v$, in relation to the potential energy curves.

From Fig.\,\ref{fig:reswf_xi}, we observe that for the DA1 process, the value of the classical turning point remains nearly constant at $R_E \approx 1.4\, a_0$ independent of the total energy $E$. Moreover, since the potentials $V^0(R)$ and $V_1^-(R)$ do not cross, it follows that $R_c\to \infty$. As a consequence, the survival factor remains practically constant, independent of the initial vibrational level and on the electron energy. From Eq.\,(\ref{eq:DAomalley}) this implies that the DA1 cross sections are determined by a constant exponential factor with superposed oscillations due to the overlap integral. For this reason, the cross sections of DA1 set, for all vibrational levels, exhibit a similar shape and remain within the same order of magnitude ranging from $10^{-6}$\,\AA$^2$ to $10^{-9}$\,\AA$^2$.

Concerning DA2 process, the crossing point is located at $R_c=3.9\,a_0$. However, defining the turning point in this case is complicated due to the shape of the resonant potential $V^-_2(R)$. Specifically, we can distinguish two regions: (i) when total energy falls within the range $V^-_2(R=8\,a_0)<E<V^0(R=8\,a_0)$ and (ii) when $E>V^0(R=8\,a_0)$. As seen in Fig.\,\ref{fig:CHxsec_X}, the cross sections exhibit two distinct behaviors: the first near the threshold up to $\approx 2\,$eV. For the DA2 process, both the exponential factor and the overlap integral play an significant and independent role in the Eq.\,(\ref{eq:DAomalley}), resulting in cross sections that vary for each vibrational level in shape and magnitude.

In the case of the DA3 process, the crossing point is located at $R_c=5.1\,a_0$. When the total energy satisfies $V^-_3(R=8\,a_0)<E<V^1(R=8\,a_0)$, the resonant width $\Gamma^3_1=0$ and the survival factor $\rho=1$. For energies $E>V^1(R=8\,a_0)$, the turning point $R_E<R_c$, the integral in Eq.\,(\ref{eq:survival}) increases slowly due to the small magnitude of $\Gamma^3_1(R)$. Consequently, the DA3 cross sections exhibit a slight exponential decay with superposed oscillations due to the overlap integral. Similarly in DA1, the set of DA3 cross sections have similar shape compacted in the range $\left[10^{-5},1\right]$\,\AA$^2$ across all vibrational levels.

Moving to the DE processes, in the LCP model, they are fully determined by the overlap integral $\langle 
\chi_c^{s} | \mathcal{V}_r | \xi^r_{s,v} \rangle$ in the Eq.\,(\ref{eq:DExsec}). 

\begin{figure}
\centering
\includegraphics[scale=.6]{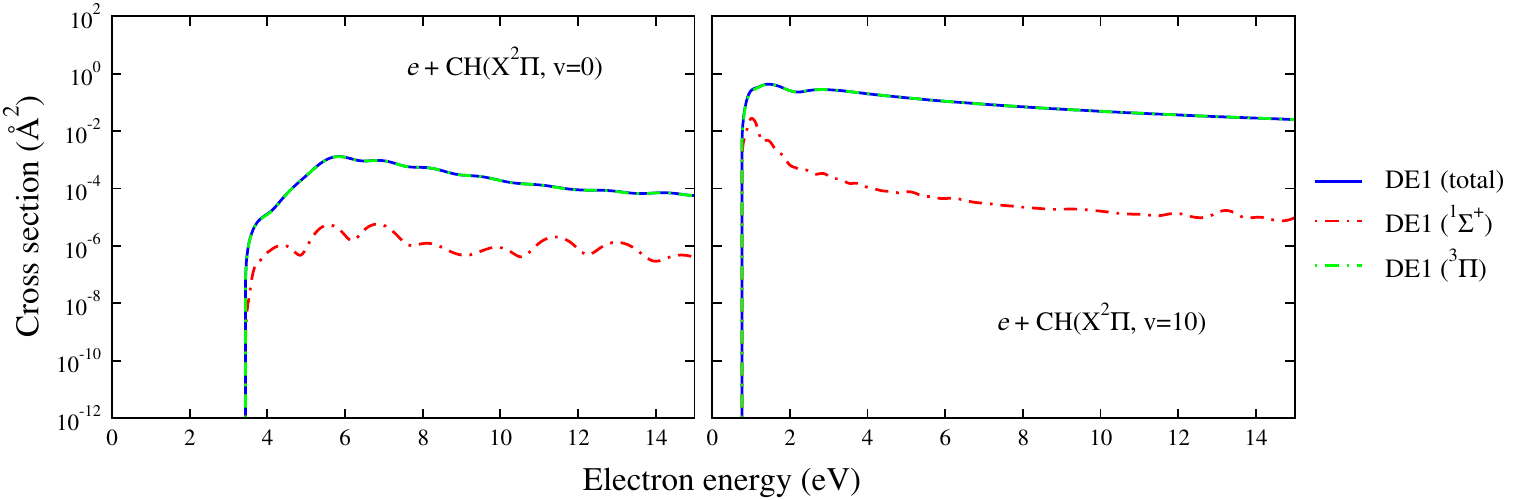} 
\caption{Cross sections of the DE1 process for the vibrational levels $v=0$ and $v=10$. The partial components due to the resonances $^1\Sigma^+$ and $^3\Pi$ are shown.  \label{fig:CHxsec_X_DE}}
\end{figure}

As reported in Table\,\ref{tab:reactions}, the reaction DE1 has two components corresponding to the resonances $^1\Sigma^+$ and $^3\Pi$. Fig.\,\ref{fig:CHxsec_X_DE} presents the total cross section along with its partial components. As can be seen, the dominant contribution comes from the $^3\Pi$ resonance.

In general, the same considerations made for DA processes can be applied to DE processes, particularly regarding the oscillatory behavior. However, the presence of certain structures near the threshold, such as in the case of DE1 for $v=5$, remains unexplained.

\begin{figure}
\centering
\includegraphics[scale=.55]{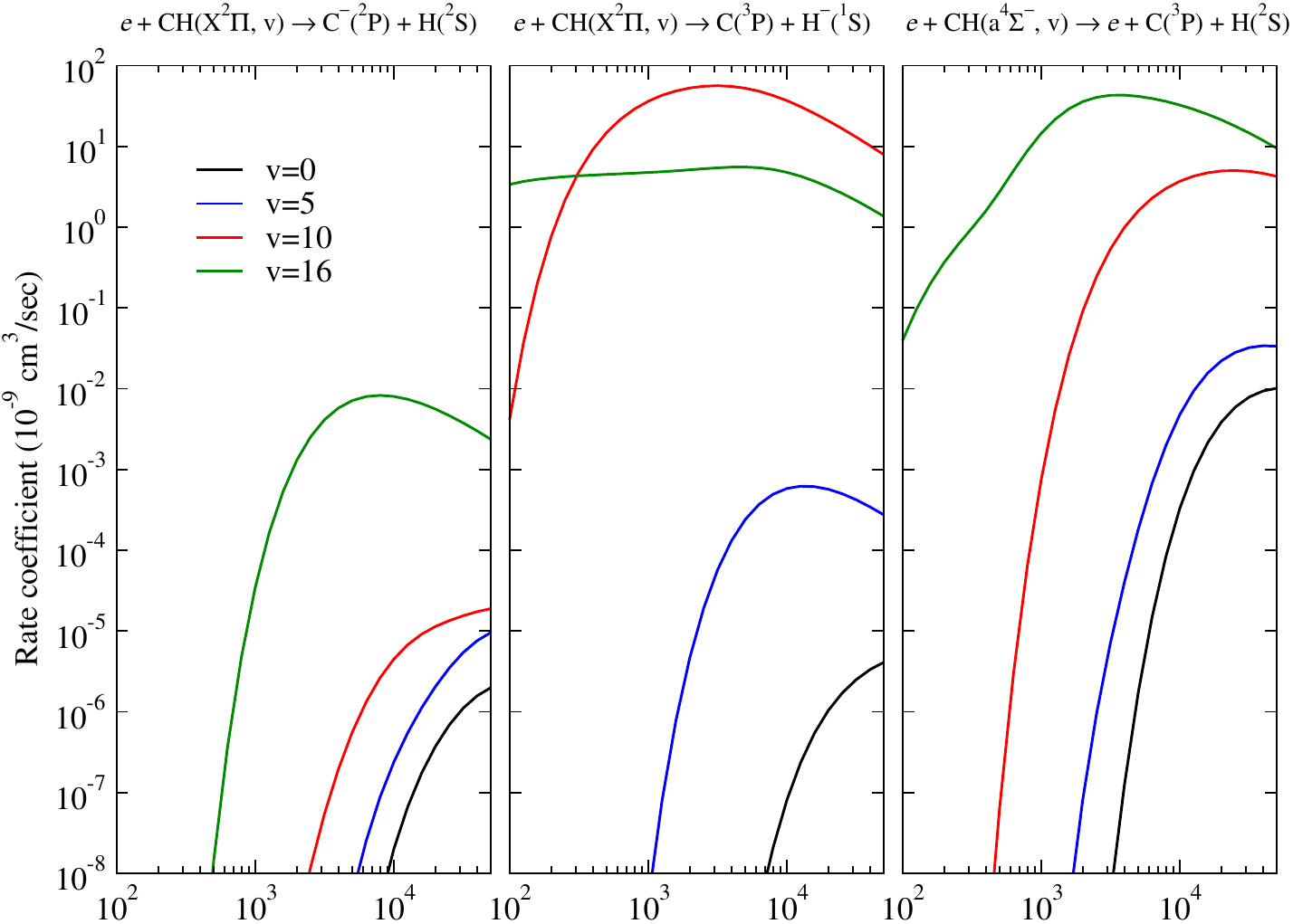}\\
\vspace{.5cm}
\includegraphics[scale=.55]{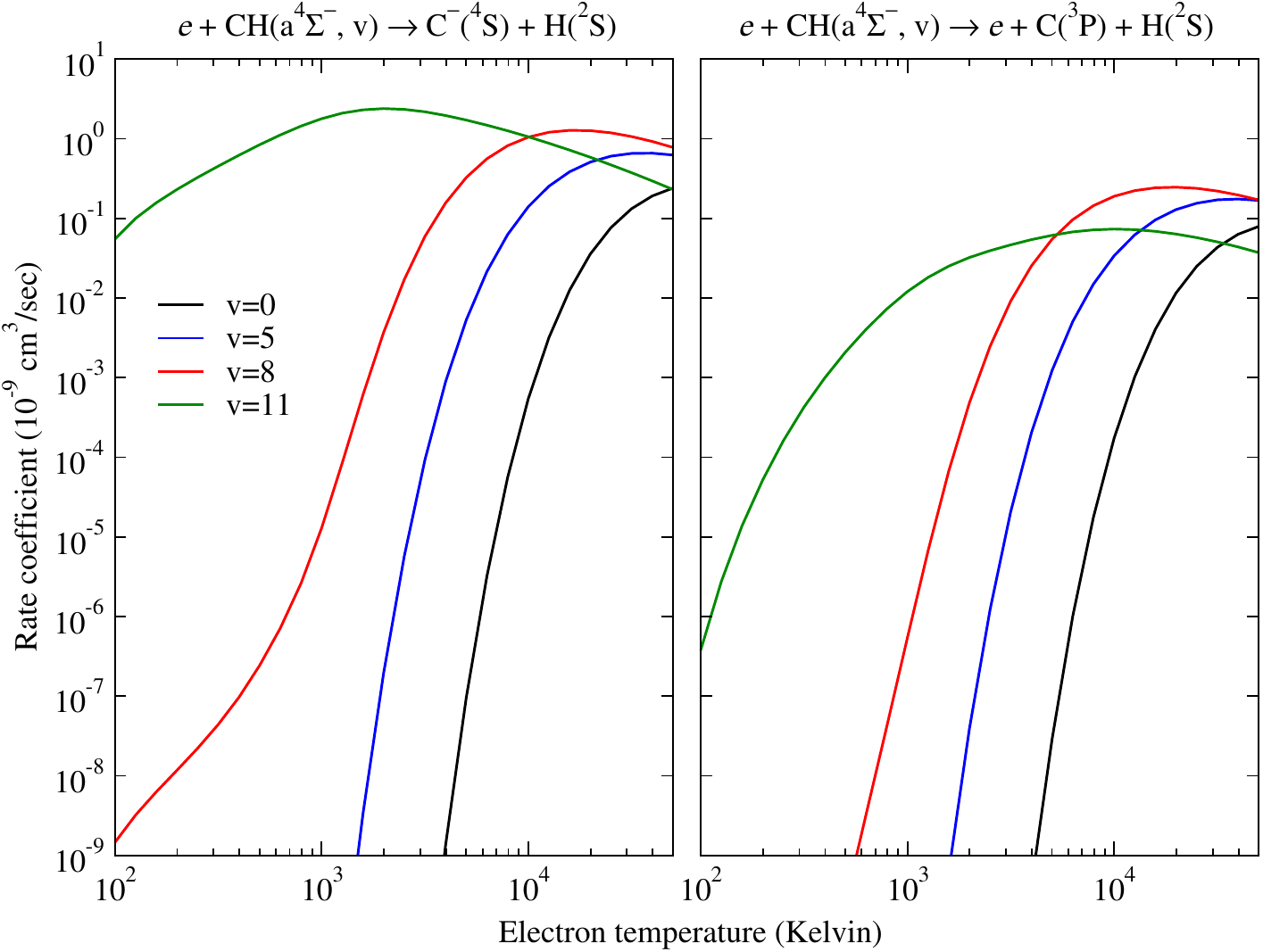} 
\caption{Rate coefficients for DA and DE processes from the ground state $\mathrm{X}\,^2\Pi$ and from the excited state $a\,^4\Sigma^-$ of CH molecule for the specified vibrational levels.  \label{fig:CHrate}}
\end{figure}

Finally, Figure\,\ref{fig:CHrate} report the Maxwellian rate coefficients defined by:
\begin{equation}
K_{s,v}(T_e) = \left( \frac{1}{m\,\pi} \right)^{1/2}\,\left( \frac{2}{k_B\,T_e} \right)^{3/2}\,\int\epsilon\,\sigma_{s,v}(\epsilon)\,e^{-\epsilon/{k_B\,T_e}}\,d\epsilon\,,
\end{equation}
for the corresponding cross sections in the Figures\,\ref{fig:CHxsec_X} and \ref{fig:CHxsec_4a} as a function of the electron temperature $T_e$.

\section{Conclusions \label{sec:conc}}

In conclusion, this paper presents a theoretical study on resonant collisions of vibrationally excited CH molecules with electrons, focusing specifically on dissociative excitation and dissociative attachment processes. This study is particularly significant, as no previous experimental data or theoretical calculations on these processes are available in the literature. Furthermore, we introduce novel electronic states for both the CH radical and its resonant anion, CH$^-$.

The scattering calculations were carried out using the local complex potential (LCP) method, while the potential energy curves (PECs) for CH and its resonant states were obtained through MRCI method using MOLPRO and \textit{ab initio} \Rmat approach. As a result, we derived vibrationally resolved cross sections and rate coefficients.

We also attempted to qualitatively explain the oscillatory structures observed in the cross sections; however, their physical origin remains not fully understood, warranting further investigation in future work.

The new cross-section and rate coefficient data contribute to a better understanding of the kinetics of non-equilibrium systems containing CH radicals, particularly regarding energy transfer between molecular vibrations and electrons, and the quenching of the system \textit{via} the formation of C$^-$ and H$^-$ anions.

These findings have important applications in plasma technologies for CO$_2$ reduction, combustion processes, and various astrophysical environments. The full dataset of cross sections and rate coefficients is available in the LXCat\,\cite{LxCat_Laporta} database.

\section*{Acknowledgements}

O.A. acknowledges hospitality from ISTP (CNR, Bari, Italy) during her visit where this work started. V.L. thanks Prof. F. Lique (Université de Rennes, France) for useful discussions and comments on the cross sections behavior and Region Bretagne for funding \textit{via} Stratégie d'Attractivité Durable (SAD) program. K.C. acknowledges support \textit{via} the DST-FIST Programme SR/FST/College/2023/1486(G).


%

\end{document}